\documentclass[doublecol,a4paper,final]{epl2}

\title{Threefold onset of vortex loops in superconductors with a magnetic core}

\author{Mauro M. Doria\inst{1,2}\thanks{E-mail: \email{mmd@if.ufrj.br}} \and Antonio R. de C. Romaguera\inst{1,2} \and M. V. Milo\v{s}evi\'{c}\inst{1,3} \and F. M. Peeters\inst{1}}
\shortauthor{M. M. Doria, A. R. de C. Romaguera et al.}

\institute{
\inst{1} Departement Fysica, Universiteit Antwerpen, Groenenborgerlaan 171, B-2020 Antwerpen, Belgium\\
\inst{2} Instituto de F\'{\i}sica, Universidade Federal do Rio de
Janeiro, C.P. 68528, 21941-972, Rio de Janeiro RJ, Brazil \\
\inst{3} Department of Physics, University of Bath, Claverton Down,
BA2 7AY Bath, United Kingdom}


\abstract{A magnetic inclusion inside a superconductor gives rise to
a fascinating complex of {\it vortex loops}. Our calculations, done
in the framework of the Ginzburg-Landau theory, reveal that {\it
loops always nucleate in triplets} around the magnetic core. In a
mesoscopic superconducting sphere, the final superconducting state
is characterized by those confined vortex loops and the ones that
eventually spring to the surface of the sphere, evolving into {\it
vortex pairs} piercing through the sample surface.}

\begin{document}

\maketitle

Examples of one- or multi-dimensional collections of topological
defects arise in various physical systems, such as vortices in
quantum fluids \cite{nem}, dislocations in solids \cite{nabarro},
global cosmic strings \cite{kib,joao}, and polymer chains
\cite{wiegel}. Changes of symmetry in their structural and dynamical
properties often lead to very rich phase transitions. In
superconductivity, the structure and symmetry of the vortex lines
has been a major issue even in the early days, when Abrikosov
\cite{abrikosov} proposed a square pattern for the lattice of
penetrating flux tubes in bulk superconductors; later on, Kleiner
{\it et al.} \cite{kra} found that the most efficient way to pack
interacting parallel rods is actually through a hexagonal pattern.
However, in systems with reduced dimensionality, the Abrikosov
symmetry of the vortex state can be strongly affected by the sample
boundaries. In mesoscopic superconductors, where the volume to
surface area ratio is small \cite{geim}, strong lateral confinement
may impose the formation of multiquanta (`giant') vortices
\cite{peeters,kanda}, while the shape of the boundary directs the
symmetry of the final vortex configuration \cite{ben}.

In this Letter, we consider a novel vortex structure that stems from
the specific {\it inhomogeneous} magnetic field of a point like
magnetic dipole with moment $\mu$ in the center of a superconducting
sphere. This magnet is the only source and sinkhole of vortex lines
and sets an axis around which the vortex state exhibits a discrete
symmetry. Because the sphere does not break this degeneracy many of
the present results also apply for a magnetic inclusion in a bulk
superconductor.

Therefore, the present analysis is very different from earlier
studies of vortex and antivortex structures in superconductor/magnet
hybrids \cite{misko1,misko2,misko3} where magnetic dots were placed
near, but {\it outside} a {\it thin} superconducting sample. While
in the latter case vortices were two-dimensional (2D) structures
characterized by a coin-like core, here they are lines in
three-dimensional (3D) space, set on a formalism that comprises the
full 3D Ginzburg-Landau theory, thus far more complex than previous
works. Consequently, novel vortex phases with profound 3D features
are found for these sets of curved vortex lines, made of {\it
confined vortex loops} (\textbf{CVL}s) and {\it external vortex
pairs} (\textbf{EVP}s).

For small $\mu$, as shown later, we find that the vortex state is
made of {\it exactly three} confined vortex loops around the dipole
axis. For an increasing magnetic moment $\mu$ more elaborate vortex
arrangements arise because of the inhomogeneity of the field. This
threefold growth of vortex loops in our system is shown here to be a
consequence of the energetic balance. Threefold symmetry has been a
source of inspiration since the sixties when its discovery in
hadronic physics has led to quarks as the basic constituents of
matter. Recently, experimental evidence was found for the onset of
trimer bound states in ultracold gases of caesium atoms
\cite{kraemer} and $Z_3$ symmetry was also proposed as a way to
solve baryon number violation in grand unified theories
\cite{agashe}.

\section{Theoretical concept} The magnetic field of the dipole
$\bm{B}=[3({\bm \mu}\cdot \hat{r})-{\bm \mu}]/r^3$ produces a mixed
state whose size can be estimated by simple arguments. Take its
magnetic field, at the equatorial plane ($\bm{\mu}\cdot \hat{r}=0$),
and compare it to the known upper and lower critical field
expressions obtained for a lattice of parallel flux lines, namely,
$H_{c2} = \Phi_0/2\pi\xi^2$, and $H_{c1} =
H_{c2}\ln\kappa/2\kappa^2$, respectively, where $\xi$ is the
coherence length and $\kappa$ is the Ginzburg-Landau constant. The
critical field regions are reached at distances
$r(H_{c2})=(\mu/\mu_0)^{1/3}\xi$ and
$r(H_{c1})=r(H_{c2})(2\kappa^2/\ln\kappa)^{1/3}$, respectively. The
condition $r(H_{c2}) \ge \xi$ holds for a magnetic moment larger
than $\mu_0=\Phi_0\xi/2\pi$. In this case the volume around the
magnetic moment, where superconductivity is destroyed has a radius
larger than the minimal length scale $\xi$, the Cooper-pair size.
The feasibility of such a magnetic domain with moment $\mu_0$ is
verified by considering a finite volume instead of a point-like
magnetic moment. Let us assume a spherically magnetized core with
radius $r_M$, and compare its magnetization $M$ to the saturation
magnetization, defined here to be one Bohr magneton per atom,
$M_0=\mu_B/(4\pi a_0^3/3)=1.49$ tesla, where $a_0\approx 0.05$
\rm{nm} is the atomic radius. The Bohr magneton is $\mu_B = \Phi_0
r_c/2\pi$, and $r_c$ is the electron classical radius. For instance,
for $\xi \approx 3.0 \; \rm{nm}$, $\mu_0$ has nearly a million
oriented Bohr magnetons ($\xi/r_c \approx 10^6$, $\xi/a_0\approx
60$). In this case the magnetized core must have radius slightly
larger than the coherence length, namely, $r_M \ge 1.7\xi$ in order
to have a magnetization smaller than the saturation value ($M \le
M_0$). Here, we consider a type II superconductor of finite size,
such that the London penetration length is much larger than the
sample and therefore, $r(H_{c1})$ is never reached. The magnetic
field streamlines around the magnetic moment are described in
spherical coordinates by $r= a\sin^2\theta$, $a$ being a parameter
associated to the loop size, and $\theta$ the angle with respect to
the magnetic moment axis. This is a special case of the so-called
rose curve, firstly described by the seventeenth century Italian
mathematician, Guido Grandi.

To investigate the superconducting state of a sample with volume
$V$, we minimize with respect to the order parameter $\psi$ the
Ginzburg-Landau (GL) free energy
\begin{eqnarray}
F = \int {{dv}\over{V}} \left( |(-i{\bf \nabla} - \frac{{\bm \mu}
\times {\bm r}}{|{\bm r}|^3})\psi|^2 - |\psi|^2 + {1 \over 2}
|\psi|^4 \right), \label{eq:glth}
\end{eqnarray}
expressed in units of the critical field energy density,
$F_0=H_c^2/4\pi$. Note that in Eq.~(\ref{eq:glth}) all distances are
scaled to $\xi$, and magnetic moment $\mu$ is given in units of
$\mu_0$. The order parameter is kept equal to zero in the center of
the sphere where the magnetic moment is located, a condition that
does not affect our results since the previously described $H_{c2}$
normal core has radius larger than $\xi$ around the dipole.  Three
different and independent numerical methods were applied to this GL
theory: simulated annealing minimization \cite{dgr}, integration of
the GL equations \cite{peeters}, and kinetic energy eigenfunction
expansion \cite{schwsaddle}. Therefore, the present results were
checked to be method independent. In what follows, we consider a
sphere with radius $R=15.0\xi$, and determine the vortex state as
the magnetic moment $\mu$ is increased.
\begin{figure}
\includegraphics[width=\linewidth]{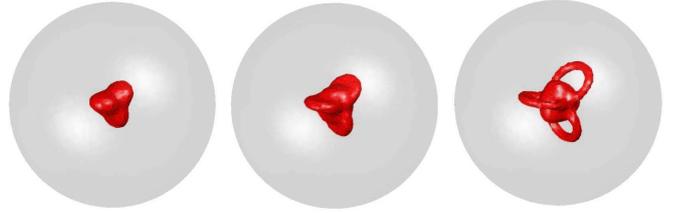}
\caption{(Colour on line) Isoplots of the Cooper-pair density for a
magnetic moment $\mu/\mu_0$ equal to $15$, $20$, and $25$ (left to
right) in a superconducting sphere of radius $15.0\xi$. Red and grey
surfaces are parts of the same isosurface drawn at 20 $\%$ of the
maximum $|\psi|^2$.}
 \label{fig1}
\end{figure}

\section{The threefold symmetry} As the magnetic dipole strongly
suppresses superconductivity inside the $H_{c2}$ core, {\it vortex
loops} nucleate as 3D objects in its vicinity, as shown in
fig.~\ref{fig1}. The following animations show a three-dimensional
view of the loop states upon rotation of the sphere for the
$\mu/\mu_0$ equal to  $19$ and $28$ states, which correspond to
embryonic and fully developed loop states, respectively:
{M19.mov}
and
{M28.mov}
(see
{multi\_media}
file for more details about the animations). Distinctively from 2D
vortices the length of the vortex line plays a fundamental role
here. We find the remarkable result - a single loop or a pair of
loops is not a stable state; three vortex loops are energetically
favorable. With increasing dipole strength {\it vortex loops
nucleate in triplets} from the $H_{c2}$ core, in a continuous
manner, i.e. through a second order transition. We should emphasize
that such a behavior was found regardless of the sample size (for
$R>10\xi$) and symmetry (e.g. sphere, cube), and is therefore a
feature that holds also for bulk samples. The threefold growth of
vortex loops is a consequence of the delicate energetic balance
among the three-dimensional curved vortex lines. We used two methods
to check its stability.
\begin{figure}
\includegraphics[width=\linewidth]{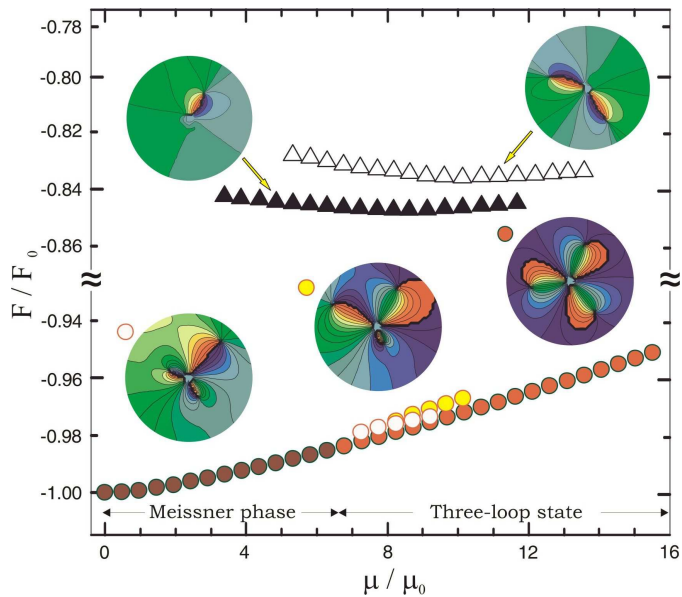}
\caption{(Colour on line) The free energy as a function of the
dipole moment, obtained by the eigenfunction expansion method.
Insets show the phase plots (blue/red - 0/2$\pi$) of the
corresponding vortex states, taken in the central cross-section
perpendicular to the dipole.} \label{fig2}
\end{figure}

The first method is the kinetic energy eigenfunction expansion
\cite{schwsaddle}, a very appropriate method to prove this because
near to the core the order parameter must be small. We expanded the
order parameter, $\Psi=\sum_{k=1}^{\Omega}C_k\psi_{k}$ in the
orthonormal eigenfunctions of the 3D kinetic energy operator
$\left(i\hbar\nabla_{3D}-e^*{\bf
A}/c\right)^2\psi_k\big/2m^*=\epsilon_k\psi_k$. In cylindrically
symmetric structures, these eigenfunctions have the form
$\psi_{k=(L,n)}=\exp(iL\phi)f_n(r,\theta)$, where $L$ is the angular
momentum, $\phi$ is the azimuthal angle around the dipole axis, and
the index $n$ counts different states with the same $L$. The
eigenfunctions $f_n$ are real and their corresponding eigenvalues
are obtained numerically for each $L$, using the Housholder
technique. Neumann boundary condition are assumed on the surface of
the sample. The typical number of considered eigenfunctions is
$\Omega$=10-50, with $L\in (-5,5)$ and $n\in (1,5)$. To search for
the free energy minimum we map the superconducting state into a 2D
cluster of $\Omega$ classical particles $(x_k, y_k) = (\Re(C_k),
\Im(C_k))$, which is governed by the energy functional obtained when
the expression for $\Psi$ is substituted in Eq.~(\ref{eq:glth}). In
other words, the energy minimization with respect to the complex
variables $C_k$ leads to the stable vortex states. Note that the
total angular momentum of the vortex loops equals zero, thus both
eigenfunctions with positive and negative $L$ must be included
[$L\in (-L_{max},L_{max})$]. Maximal considered vorticity $L_{max}$
determines the maximal achievable number of vortex loops. For
$L_{max}=1$ and $L_{max}=2$, we were able to stabilize one and two
loops respectively, but their energy was always significantly higher
than the Meissner state, as shown in fig.~\ref{fig2}. The encircling
discontinuity of $2\pi$ (blue to red/red to blue) in the phase
contour plots of fig.~\ref{fig2} signals a vortex/antivortex.
However, when eigenfunctions with higher angular momenta were
included in the calculation, the energy minima associated with just
one and two loops ceased to exist. Instead, stable states with one
pronounced loop and two `satellite' ones (and vice versa) were
found, yet with higher energy than the state with perfect threefold
symmetry.

\begin{figure}
\includegraphics[width=1.0\linewidth]{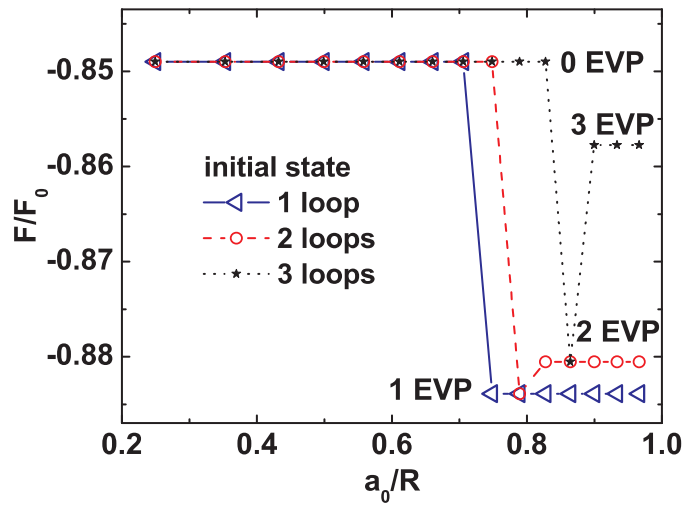}
\caption{(Colour on line) The free energy (arbitrary units) of the
final relaxed state is shown  versus the ratio loop size to radius,
$a_0/R$, for the initial state. The magnetic moment is equal to
$\mu/\mu_0=25$ and the initial states are made up of either one, two
or three symmetrically arranged loops. If the initial state is near
to the edge, the onset of \emph{external vortex pairs} becomes
possible otherwise not.} \label{fig3}
\end{figure}

\begin{figure}
\includegraphics[width=\linewidth]{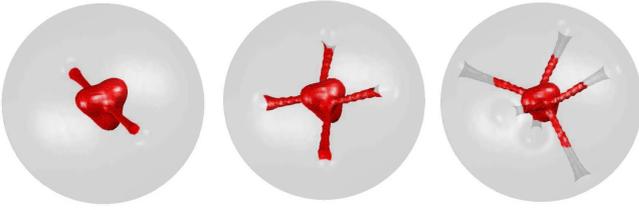}
\caption{(Colour on line) Isoplots of the Cooper-pair density for
three $\mu/\mu_0=25$ accessible states, namely, 1, 2 and 3 \emph{
external vortex pairs}  (left to right) in a superconducting sphere
of radius $15.0\xi$. Red and grey surfaces are parts of the same
isosurface drawn at 20 $\%$ of the maximum.} \label{fig4}
\end{figure}

The second method used to support the stability of the three loop
state is based on the minimization of the Ginzburg-Landau free
energy through the simulated annealing procedure \cite{dgr} starting
from an initial constructed state, $\Psi_N(a,\phi)$, which depends
on the following variables: the loop size, $a$, the azimuthal angle
around the dipole axis, $\phi$, and the number of zeros of this wave
function, $N$. This initial state is built such that the condition
$\Psi_N(a_0,\phi_0)=0$ holds for a single loop position, $a=a_0$,
and for a special azimuthal configuration determined by the
condition $\exp(i N \phi_0)=1$. Hence this initial state vanishes
along $N$ symmetric loops. Obviously the above requirements don't
uniquely define $\Psi_N(a,\phi)$, but we don't discuss the specific
wavefunction taken here as we found invariance in our results with
respect to this choice. For a fixed value of $\mu$ the numerical
search for the minimum is carried and the final state free energy is
shown in fig.~\ref{fig3} versus the initial loop size, $a_0/R$.
Three distinct initial states were taken, i.e, $N=1$, $2$ and $3$
CVLs. Indeed the numerical procedure yielded just one single final
state with $3$ CVLs regardless of the number of CVLs in the initial
state, thus proving its stability. However this holds only for a
small $a_0/R$ ratio, according to fig.~\ref{fig3}. For a large
$a_0/R$ ratio several final states are reached. The proximity to the
boundaries explains this puzzling fact that for $a_0/R > 0.73$ the
free energy minimum depends on the initial state, as shown in
fig.~\ref{fig3}. The proximity of $N$ loops to the surface of the
sphere causes the onset of EVPs during the annealing procedure. In
fact for the three highest values of $a_0/R$, the number of EVPs is
exactly the number of CVLs in the initial state ($N$). This striking
correlation leaves no doubt that an initial state with a CVL very
near to the edge most easily turns into an EVP during the
minimization procedure. Regardless of the number of EVP all the
observed final vortex states of fig.~\ref{fig3} display three
embryonic confined loops very near to the $H_{c2}$ core, as shown in
fig.~\ref{fig4}. The stability of states with other than three loops
is only possible by explicitly breaking the azimuthal symmetry, such
as in case of a rigid lattice of magnetic moments inside a bulk
superconductor \cite{doria04a,doria04b}.

%
%
Thus for a small superconducting sphere, i.e., with a ratio $R/\xi$
not very large, the presence of boundaries strongly influences the
vortex state as seen in figs. \ref{fig3} and \ref{fig4}. This is
because  CVLs very near to the superconductor's external surface
eventually break and spring to the surface in the form of EVPs. Next
we discuss some general aspects of these states. The switch of a CVL
into an EVP induces a first order transition. EVPs are shown in
fig.~\ref{fig4} in a grey and red color scheme. Red and grey colors
describe parts of a single isosurface contained in distinct
concentric volumes. Inside a inner cubic cell the isosurface is red
and in the remaining volume grey. Thus a CVL is always red, whereas
the EVPs are double colored, red and grey. The external isosurface,
including the regions of contact with the EVPs are represented in
grey color. Thus for the finite size superconductor the vortex state
is generally classified according to the number of CVLs and of EVPs,
an important difference to previous 2D studies
\cite{misko1,misko2,misko3} where the only distinction is between
`giant' and `multi' vortex states. States with multiple EVPs
manifest as a collection of vortices on the sample surface, but are
conjoined to a giant vortex in the sample center. The total angular
momentum (`winding number', or `vorticity') increases by one for
each additional EVP. According to fig.~\ref{fig4} this giant vortex
also contains three embryonic CVLs. The following animations show
three-dimensional views of some $\mu/\mu_0$ states, namely, for
$24$, $34$, and $31$. These states show 3 embryonic CVLs with 1, 2
and 3 EVPs, respectively: M24.mov, M34.mov and M31.mov (see
multi\_media.doc file for more details about the animations).
\begin{figure}[t]
\includegraphics[width=\linewidth]{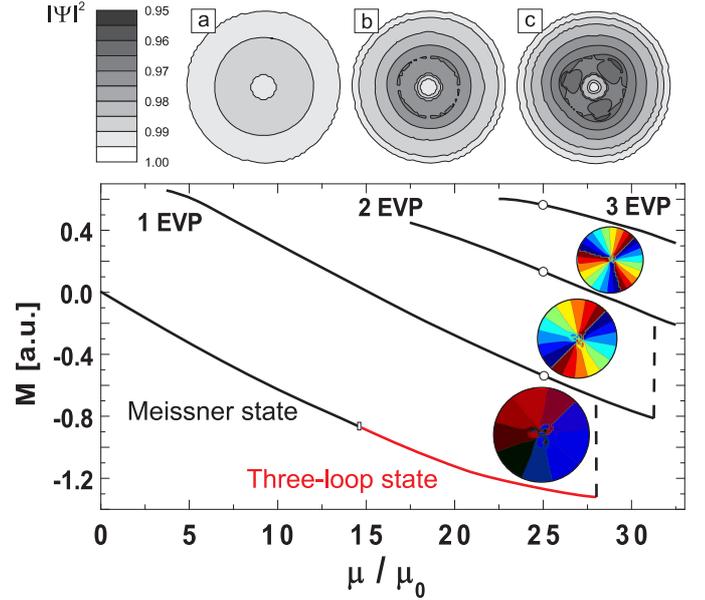}
\caption{(Colour on line) Sample magnetization (in arbitrary units)
vs. dipole moment. Insets show the phase in the equatorial plane of
the sample for the 1-3 \emph{external vortex pair} states of
fig.~\ref{fig4} (for $\mu/\mu_0=25$, open dots). (a)-(c) $|\psi|^2$
contourplots of the top half hemisphere of the sample, for the three
\emph{confined vortex loops} state and $\mu/\mu_0=20$, $24$, and
$28$, respectively.} \label{fig5}
\end{figure}
The CVL to EVP (first order) transitions lead to jumps in the
magnetization ($M=-\partial F/\partial \mu$), shown in
fig.~\ref{fig5} as a function of the dipole moment $\mu$. Note that
this feature directly leads to accessible experimental detection of
EVP states, e.g. by Hall-probe magnetization measurements. However,
the observation of CVL states is more demanding and requires a
combination of various complementary microscopic experimental
techniques, such as Small Angle Neutron Scattering (SANS), muon spin
rotation ($\mu$SR) \cite{musr}, and possibly scanning tunneling
microscopy on curved surfaces to detect the threefold slight
suppression of the order parameter on the surface of the sphere [as
shown in grey scale of figs.~\ref{fig5}(a-c)].
\begin{figure}[t]
\includegraphics[width=\linewidth]{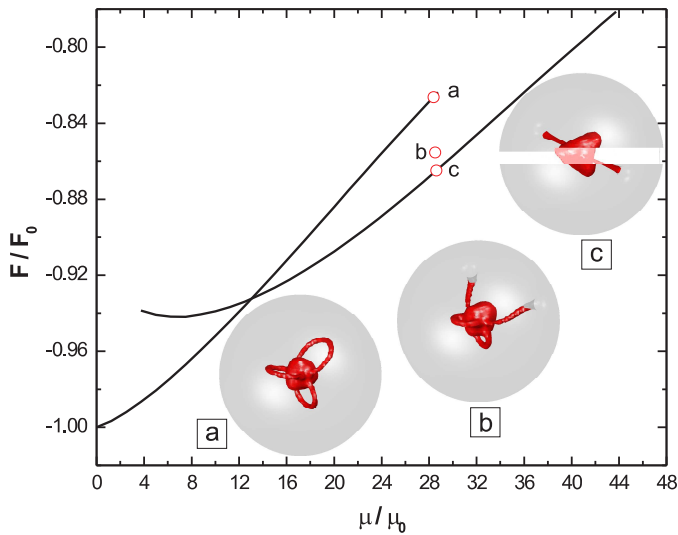}
\caption{(Colour on line) The free energy curves versus dipole
moment for 0 and 1 \emph{external vortex pair} state obtained by the
simulated annealing method. The \emph{confined vortex loop} state is
metastable for $\mu/\mu_0>12.5$ and becomes unstable at
$\mu/\mu_0=28.5$. The isosurfaces in the insets (a-c) show snapshots
of the consequent 0 to 1 \emph{external vortex pair} transition.}
\label{fig6}
\end{figure}

\section{The evolution of the vortex state} The onset of EVPs
unveils interesting aspects of the threefold symmetry because of the
soft breaking interactions that drive such first order transitions.
A CVL becomes unstable near to the surface of the superconductor and
eventually breaks apart giving rise to a EVP, as shown in figs.
\ref{fig2}-\ref{fig5}. For our spherical sample the three initially
equivalent loops are naively expected to grow and reach the surface
simultaneously, turning into three EVPs. Nevertheless one and two
EVPs are also observed implying that the threefold symmetry is
disturbed during its evolution by some soft breaking mechanism, here
caused by the use of a cubic grid. The evolution of the remaining
two CVL, as the third one turns into an EVP, is quite remarkable.
The outgrowth of one particular CVL from the threefold state causes:
(i) the shrinkage of the other two ones and (ii) the appearance of a
new CVL to recover the threefold symmetry near the $H_{c2}$ core.
This shows that it is easier to recover the threefold symmetry near
the $H_{c2}$ core because there the CVLs are less energetically
costly to produce. The development of the symmetry during the
simulated annealing transition from $0$ to $1$ EVP is shown in
fig.~\ref{fig6}. A 3 CVL state becomes unstable at $\mu/\mu_0=28.5$
[fig.~\ref{fig6}(a)] and evolves, under controlled annealing
temperature, to a 1 EVP state. Threefold symmetry is observed {\it
both below and above} the transition point. A trapped intermediate
(unstable) configuration during this transition is shown as inset
(b) of fig.~\ref{fig6}. We confirmed these results by a saddle-point
analysis \cite{schwsaddle}, carried through the kinetic energy
eigenfunction expansion. Moreover we have shown that the presence of
an external current applied along the magnetic dipole can induce
this transition, as shown in the following animation done for
$\mu/\mu_0=10$:
{ZERO-TO-ONE.mov}
(see
{multi\_media}
file for more details about the animations). The density increases
away from the $H_{c2}$ core at the center but eventually this growth
stops and the trend reverts to a decreasing pattern  near to the
surface of the mesoscopic sphere. Outside the order parameter must
rigourously vanish, and so, there is always an external isosurface
density near to the surface of the mesoscopic sphere with constant
and small $|\psi|^2$. The figs.~\ref{fig1},~\ref{fig4}
and~\ref{fig6} show it in grey scale as part of the isosurface drawn
at 20 $\%$ of the maximum $|\psi|^2$.
\section{Conclusion}
Using several independent theoretical approaches, we found a {\it
threefold} embryonic growth of vortex loops from the $H_{c2}$ core
surrounding the magnetic moment in a superconducting sphere. As
shown here, the superconducting state with three vortex loops is
energetically favorable over one or two vortex loops regardless of
the sample boundaries. Therefore, our results also apply to bulk
samples. We expect that the present scenario of loop breaking at the
sample surface can be useful to understand properties of a bulk
superconductor with some distribution of magnetic domains in its
interior, where growing vortex loops may interconnect neighboring
magnets rather than spring to the surface, which would lead to the
experimentally observable spontaneous vortex phase \cite{ng}. The
puzzling properties of the recently discovered ferromagnetic
superconductors \cite{bluhm} and superconducting ferromagnets
\cite{sakai,jorgensen} are indicative of vortices stemming from an
internal magnetic field. We expect that the recent advancements in
material synthesis, which have led to several fabricated
nano-composites, such as the MgB$_2$ superconductor with embedded
magnetic Fe$_2$O$_3$ nanoparticles \cite{snezhko}, or Gd particles
incorporated in a Nb matrix \cite{palau}, can be used to obtain the
predicted vortex patterns made of confined vortex loops and external
vortex pairs.

\acknowledgments A. R. de C. Romaguera acknowledges support from
CNPq (Brazil). M. M. Doria acknowledges support from CNPq (Brazil),
FAPERJ (Brazil), the Instituto do Mil\^enio de Nanotecnologia
(Brazil) and BOF/UA (Belgium). M. V. Milo\v{s}evi\'{c} and F. M.
Peeters acknowledge support from the Flemish Science Foundation
(FWO-Vl), the Belgian Science Policy (IUAP) and the ESF-AQDJJ
network. M. V. Milo\v{s}evi\'{c} is currently a Marie-Curie Fellow
at University of Bath, UK.

\end{document}